\documentclass[reprint,aps,prd,twocolumn,notitlepage,nofootinbib,
showpacs,preprintnumbers,superscriptaddress]{revtex4-1}
\usepackage[T1]{fontenc}
\usepackage[utf8]{inputenc}
\usepackage{bm}
\usepackage{amsmath}
\usepackage{amssymb}
\usepackage{esint}
\usepackage{color}
\usepackage{graphicx}
\PassOptionsToPackage{normalem}{ulem}
\usepackage{ulem}

\usepackage{subcaption}
\usepackage{graphicx}

\usepackage{times}
\usepackage{hyperref}
\usepackage{amsthm,bm,mathrsfs,stmaryrd,amsmath}

\newcommand{\be}{\begin{equation}}
\newcommand{\ee}{\end{equation}}
\newcommand{\ba}{\begin{eqnarray}}
\newcommand{\ea}{\end{eqnarray}}

\newcommand{\nn}{\nonumber\\}

\def\a{\alpha}
\def\b{\beta}

\def\G{\Gamma}
\def\d{\delta}

\def\l{\lambda}

\def\p{\pi}

\def\be{\begin{eqnarray}}
\def\ee{\end{eqnarray}}

\def\a{\alpha}
\def\b{\beta}

\def\l{\lambda}

\def\G{\Gamma}
\def\d{\delta}
\def\nn{\nonumber\\}

\newcommand\<\langle
\renewcommand\>\rangle

\begin{document}

\title{Principle of minimal singularity for Green's functions}

\author{Wenliang Li}
\email{liwliang3@mail.sysu.edu.cn}
\affiliation{School of Physics, Sun Yat-Sen University, Guangzhou 510275, China}

\begin{abstract}
Analytic continuations of integer-valued parameters can lead to profound insights, such as  
angular momentum in Regge theory, 
the number of replicas in spin glasses, 
the number of internal degrees of freedom,
the spacetime dimension in dimensional regularization and Wilson's renormalization group. 
In this work, we consider a new kind of analytic continuation of correlation functions,  inspired by two recent approaches to underdetermined Dyson-Schwinger equations in $D$-dimensional spacetime. 
If the Green's functions $G_n=\<\phi^n\>$ admit analytic continuation to complex values of $n$, 
the two different approaches are unified by a novel principle 
for self-consistent problems:   
Singularities in the complex plane should be minimal. 
This principle manifests as the merging of different branches of Green's functions 
in the quartic theories. 
For $D=0$, we obtain the closed-form solutions of the general $g\phi^m$ theories, 
including the cases with complex coupling constant $g$ or non-integer power $m$.   
For $D=1$, we derive rapidly convergent results for the Hermitian quartic and non-Hermitian cubic theories by minimizing the complexity of the singularity at $n=\infty$. 
\end{abstract}

\maketitle
\flushbottom

\section{Introduction} 
Analytic continuations of real-valued parameters are fascinating. 
Two early examples are the imaginary magnetic field in the Yang-Lee theory of phase transitions \cite{Yang:1952be, Lee:1952ig} and 
the time coordinate in Wick rotation that connects quantum mechanics to statistical mechanics \cite{Wick:1954eu}. 
The complexification of discrete parameters is even more intriguing, 
such as complex angular momentum in Regge theory of scattering \cite{Regge:1959mz} and 
Brout's replica trick for spin glasses \cite{Brout:1959zz}. 
Furthermore, a non-integer number of degrees of freedom can unify different physical systems.  
For instance,  
de Gennes's mapping relates the $n\rightarrow 0$ limit of 
the $n$-vector model to self-avoiding polymers \cite{deGennes:1972zz}, 
while Fortuin-Kasteleyn's random cluster model connects the $q\rightarrow 1$ limit of 
the $q$-state Potts model with the percolation problem \cite{Fortuin:1971dw}. 
The spacetime  dimension can also be continued, 
leading to the method of dimensional regularization \cite{Bollini:1972bi,tHooft:1972tcz} 
and Wilson-Fisher fixed points of the renormalization group flow \cite{Wilson:1971dc}. 

In this work, we consider a new kind of analytic continuation associated with Green's functions $G_n$.  
In quantum field theory, Green's functions describe the correlation of quantum fields and encode information about the particle spectrum and vacuum structure.  
As quantum equations of motion, 
the Dyson-Schwinger (DS) equations \cite{Dyson:1949ha,Schwinger:1951ex,Schwinger:1951hq} 
imply that the Green's functions are related to each other, 
providing a self-consistent way to determine them nonperturbatively.   
However, the nonperturbative DS equations usually form an underdetermined system \cite{Bender:1988bp}.  
Additional assumptions or constraints are needed to close the system. 
A simple scheme is to set the higher connected Green's functions to zero. 
Even though one may find convergent results, 
their limiting values deviate from the exact values, as emphasized recently in \cite{Bender:2022eze,Bender:2023ttu}. 
To resolve this issue, Bender {\it et al.} replaced the naive vanishing constraints with more sophisticated approximations from the large $n$ asymptotics \cite{Bender:2022eze,Bender:2023ttu}.  
Alternatively, one can resolve the indeterminacy unbiasedly using the null state condition \cite{Li:2023nip}. 
For unitary solutions, one can also use positivity to constrain the solution space
\cite{Anderson:2016rcw,Lin:2020mme,Kazakov:2021lel,Kazakov:2022xuh}. 
\footnote{Positivity constraints were introduced to the bootstrap studies of conformal field theory in \cite{Rattazzi:2008pe}. }

The use of large $n$ asymptotic behaviors in \cite{Bender:2022eze,Bender:2023ttu} was  
based on an implicit assumption. 
To study different Green's functions at the same time, 
the full set of $G_n=\<\phi^n\>$ should exhibit good analytic behavior in $n$, 
but $n$ is usually thought of as an integer parameter. 
This is reminiscent of the angular momentum quantum number, 
which is usually known as a quantized parameter that takes discrete values. 
In 1959, Regge analytically continued the angular momentum to complex values, 
leading to deep insights into the asymptotic behavior of scattering amplitudes \cite{Regge:1959mz}. 
This original idea had significantly influenced the developments of the bootstrap program \cite{Chew:book, Chew:1961ev}. 
As the DS equations furnish a self-consistent method for determining Green's functions, 
the analytic continuation in $n$ should also provide useful insights. 

There is another motivation for complexifying $n$.
Inspired by the properties of Yang-Lee edge singularities\cite{Yang:1952be, Lee:1952ig,Kortman:1971zz,Fisher:1978pf,Cardy:1985yy}, 
Bessis and Zinn-Justin studied the quantum mechanical version of the $i\phi^3$ theory 
and noticed the reality of the energy spectrum despite non-Hermiticity. 
Later, Bender and Boettcher proposed a novel family of non-Hermitian $\mathcal {PT}$-invariant theories with real and bounded spectra \cite{Bender:1998ke}, 
in which the power of the interaction term can take non-integer values. 
To bootstrap the cases with non-integer power, 
it is inevitable to take into account $\<\phi^n\>$ with non-integer $n$ 
\cite{Li:2022prn}. 

Along these lines, we revisit the indeterminacy issue of the self-consistent equations 
from the complexified-$n$ perspective. 
It turns out that there exists a novel principle that can resolve the problem: 
{\it Singularities in the complex plane should be minimal.} 
\footnote{More precisely, 
the one-point compactification of the set of complex numbers with the point at infinity 
is called the extended complex plane. } 
The null state approach \cite{Li:2023nip} mentioned above can be viewed as 
an unbiased way to minimize the complexity of singularities. 

Below, we will use this novel principle to determine the Green's functions $G_n$,  
including the real and complex solutions, 
such as the non-Hermitian cases with $\mathcal {PT}$ symmetry \cite{Bender:2022eze,Bender:2023ttu, Bender:1998ke, Bender:1999ek, Bender:2007nj,Bender:2010hf,r5}. 
In quartic theories, the principle of minimal singularity manifests as 
the merging of different branches of Green's functions. 

\section{Zero-dimensional theories} 
Let us consider the $D=0$ theory with a monomial potential.  
The Lagrangian is
\be
\mathcal L[\phi]=g \phi^m\,,
\ee
which is the $D=0$ version of multi-critical models.  
We will present the closed-form solutions for generic $m$ and $g$. 
Here the power $m$ is not necessarily a positive integer and 
the coupling constant $g$ can be a complex number. 

\subsection{Dyson-Schwinger equations}
We want to determine the Green's functions using the DS equations. 
The Green's functions can be obtained from the generating function $Z[J]$:
\be
G_n=\<\phi^n\>_{J=0}=\frac 1 {Z[0]}\frac{\d^n Z[J]}{\d J^n}\Big|_{J\rightarrow 0}\,,
\ee
where
$
Z[J]=\int d\phi\,e^{-\mathcal L[\phi]+J\phi}
$ is an ordinary integral and the path of integration depends on the choices of Stokes sectors. 
An infinitesimal change of the integration variable $\phi$ leads to the quantum equation of motion
\be
\<d\mathcal L/d\phi\>_{J}=gm\<\phi^{m-1}\>_{J}=\<J\>_{J}\,.
\ee
The $J\rightarrow 0$ limit gives $G_{m-1}=0$.
If we take some $J$ derivatives before setting $J=0$, 
we obtain the DS equations:
\be
\label{0D-DS-general}
G_{n+m}=g^{-1}\left(\frac {n+1}m\right)G_n\,. 
\ee
The normalization convention implies 
$G_0=1$.  
The general solutions for the Green's functions are given by
\be
\label{sol-0D}
G_{n'+pm}=g^{-p} \left(\frac {n'+1}m\right)_p G_{n'}\,,
\ee
where $0\leq n'<m$, and $(a)_b=\G(a+b)/\G(a)$ is the Pochhammer symbol. 
Usually, $n'$ and $p$ are integers. 
For the moment, we assume that  $m$ is an integer and $m\geq 3$, 
so the system is underdetermined.  
In terms of exponential functions, 
we find a general-$n$ expression:
\be
\label{sol-0D-exp}
G_n=\left(\sum_{k=0}^{m-1}\,c_k\,e^{2\pi i\frac {kn}{m}}\right)g^{-\frac{n}{m}} 
\left(\frac {1}m\right)_{n/m}\,,
\ee
which solves the DS equations \eqref{0D-DS-general} at both integer and non-integer $n$. 
The coefficients $c_k$ are linear in the free parameters $(G_1,\dots, G_{m-2})$.  
In general, the essential singularity at $n=\infty$ is   
a superposition of $m$ types of singular behaviors. 

\subsection{Principle of minimal singularity}
According to some explicit examples at integer $m$, we notice that the exact solutions 
take the specific form
\be
\label{0D-general-sol}
G_n=\frac{1-\a^{n+1}}{1-\a}\, \b^{n}\,g^{-\frac{n}{m}} 
\left(\frac {1}m\right)_{n/m}\,,
\ee
where $\a$ and $\b$ satisfy the periodicity conditions
\be
\label{periodicity-general}
\a^m=1\,,\quad
\b^m=1\,,
\ee
and the non-degeneracy condition 
\be
\label{non-degeneracy-general}
\a\neq 1\,.
\ee 
The $\a$ part is related to $G_{m-1}=0$ and encodes the relative singular behavior. 
The $\b$ part is a $D=0$ analog of the Symanzik/Sibuya rotation \cite{Sibuya}. 

The exact solutions are labeled by two integers. 
Since $\a$ is associated with the relative singular behavior, 
we can select the independent solutions by $\text{Im}(\a)\geq 0$. 
The number of possible types of singular behaviors increases with $m$. 
To show that \eqref{0D-general-sol}, \eqref{periodicity-general}, \eqref{non-degeneracy-general} indeed encode the exact solutions, 
let us examine the explicit examples of $m=3,4$. 

For the cubic theory with $m=3$, 
the two choices of $\a$ are not independent and we choose 
$\a=e^{2\pi i\frac {1} 3}$, so we have
\be
G^{(m=3)}_1=-e^{2\pi i\frac {k-1} 3} {g^{-\frac 1 3}} \frac{\G(2/3)}{\G(1/3)}\,,\quad
\ee
where $k=0,1,2$. 
We obtain the three exact solutions
\be
\label{D0-cubic-exact-sol}
G^{(m=3)}_n=\frac{1-e^{2\pi i\frac{n+1}{3}}}{1-e^{2\pi i\frac{1}{3}}}
\frac {e^{2\pi i\frac{kn}{3}}} {g^{n/3}}  \left(\frac {1}3\right)_{n/3}\,.
\ee
For $g=i/3$, the $\mathcal {PT}$-symmetric case corresponds to $k=2$. 

In the quartic case with $m=4$, the independent choices are 
\be
\a^{(m=4)}=e^{2\pi i\frac 1 4}\,,\quad e^{2\pi i\frac 2 4}\,.
\ee
They give all the exact solutions: 
\begin{itemize}
\item
In the first case $\a=i$, there are four solutions: 
\be
G^{(m=4)}_n=\frac{1-i^{n+1}}{1-i}\, \frac{e^{2\pi i\frac {kn} 4}}{g^{n/4}} \left(\frac {1}4\right)_{n/4}\,,
\ee
where $k=0,1,2,3$. 
For $g=-\frac 1 4$, 
the $\mathcal {PT}$-symmetric solution is associated with $k=3$.
\item
In the second case $\a=-1$, all the parity-odd Green's functions vanish.
We have two solutions:
\be
\label{D0-quartic-exact-sol}
G^{(m=4)}_n=\frac{1+(-1)^n}{2}\,\frac{(\pm 1)^{n/2}}{ g^{n/4}} \left(\frac {1}4\right)_{n/4}\,.
\ee
For $g>0$, the standard Hermitian solution corresponds to the positive case with $(+1)$. 
\end{itemize} 

We verify that the exact solutions at other concrete $m$ are also given by \eqref{0D-general-sol}, \eqref{periodicity-general}, \eqref{non-degeneracy-general}. 
For instance, the non-Hermitian quintic case with $m=5$ and $g=-i/5$ has two $\mathcal {PT}$ symmetric solutions at 
$G_1\approx -1.08i,\,0.41i$. They correspond to 
$(\a,\b)=(e^{2\pi i\frac 1 5}\,,\, e^{2\pi i\frac 3 5})\,,\,
(e^{2\pi i\frac 2 5},\, 1)$\,. 

What distinguish the exact results \eqref{0D-general-sol}, \eqref{periodicity-general}, \eqref{non-degeneracy-general} from other self-consistent solutions in \eqref{sol-0D-exp}? 
We notice that the exact solutions have only two types of singular behaviors at $n=\infty$, 
rather than all the $m$ types in \eqref{sol-0D-exp}.  
Since we need at least two terms to be compactible with $G_{m-1}=0$, 
the exact solutions are minimally singular. 
Therefore, we are led to introduce the principle of minimal singularity. 

Let us use the principle of minimal singularity to derive \eqref{0D-general-sol}, \eqref{periodicity-general}, \eqref{non-degeneracy-general}. 
According to this novel principle, 
we should minimize the complexity of the singularity structure. 
The results are the solutions with two types of singular behaviors at $n=\infty$, 
i.e., only two coefficients in \eqref{sol-0D-exp} are non-zero. 
Suppose that they are labelled by $k_1$ and $k_2$. 
Then the constraint $G_{m-1}=0$ fixes their relative coefficient. 
The $k$ summation becomes $c_{k_2}(1-e^{2\pi i\frac {k_1-k_2}{m}(n+1)})e^{2\pi i\frac {k_2 n}{m}}$. 
The remaining coefficient $c_{k_2}$ is determined by $G_0=1$. 
In this way, we obtain the closed-form results \eqref{0D-general-sol}, \eqref{periodicity-general}, \eqref{non-degeneracy-general}.
\footnote{We focus on the principal values of $k$,  
which satisfy $0\leq k <m$. 
Other choices can lead to different results at non-integer $n$,  
but they are redundant at integer $n$ due to $e^{2\pi i\frac {m}{m}n}=1$. }

In analogy to the multi-cut solutions in matrix models 
(see \cite{Lin:2020mme,Kazakov:2021lel} for summaries and references therein), 
one can study more complicated solutions by 
keeping more $c_k$ in \eqref{sol-0D-exp}.  
For example, a next-to-minimal solution can have three independent exponential terms
\be
G_n=\left(\sum_{j=1,2,3}\,c_{k_j}\,e^{2\pi i\frac {k_j n}{m}}\right)g^{-\frac{n}{m}} 
\left(\frac {1}m\right)_{n/m}\,,
\ee  
then there remains one free parameter because $G_{m-1}=0$ and $G_{0}=1$ do not fix all the  coefficients $c_{k_j}$. 
The simpler solutions are interpolated by the more complicated ones. 
The minimally singular solutions are the starting point for a more thorough classification 
based on reducibility of singularities. 
\footnote{In the sense of theory space, the fixed points of renormalization group flow and the corresponding conformal field theories can be viewed as minimally singular quantum field theories. } 

\subsection{Large $n$ expansion}
We can also derive the exact solutions using the large-$n$ expansion. 
Let us consider the cubic example for simplicity. 
As the $\a$ part is responsible for the vanishing Green's functions at $n=3p+2$, 
we will focus on the remaining part of \eqref{D0-cubic-exact-sol}. 
For $m=3$, the DS equations \eqref{0D-DS-general} give a recursion relation between $G_{n}$ and $G_{n+3}$. We have 
\be
\label{0D-DS-cubic}
m=3:\quad G_{n+3}=g^{-1}\left(\frac {n+1}3\right)\,G_n\,.
\ee 
We assume the existence of a stronger relation
\be
G_{n+1}=F_n\,G_{n}\,,
\ee
where $F_n$ should be minimally singular in $n$. 
We have
\be
\label{0D-DS-cubic-f}
F_n\,F_{n+1}\,F_{n+2}=g^{-1}\frac {n+1}{3}\,,
\ee
which can be systematically solved at large $n$. The leading terms in $1/n$ are
\be
F_{n}=\frac {e^{2\pi i\frac {k}{3}}} {g^{1/3}}\,\frac{n^{1/3}}{3^{1/3}}
\left(1+\frac 1 {9 n^{2}}-\frac {13}{{81}n^{4}}
+\dots\right)\,,
\ee
where $k=0,1,2$. 
In general, the resummation of the $1/n$ series is not unique. 
To minimize the complexity of singularities, we use the gamma function to construct the ``cube root''  of the DS equations \eqref{0D-DS-cubic}, \eqref{0D-DS-cubic-f}:
\be
F_n=\frac {e^{2\pi i\frac {k}{3}}} {g^{1/3}}\left(\frac{n+1}{3}\right)_{1/3}\,, 
\ee
then we obtain the exact solutions in \eqref{D0-cubic-exact-sol}. 
The large-$n$ expansion is useful 
if the general solutions are not expressed as known functions. 
The effectiveness of the leading asymptotic behavior had been shown in \cite{Bender:2022eze,Bender:2023ttu} for the connected Green's functions.  

\subsection{Non-integer power $m$}
\begin{figure}[h]
	\centering
		\includegraphics[width=1\linewidth]{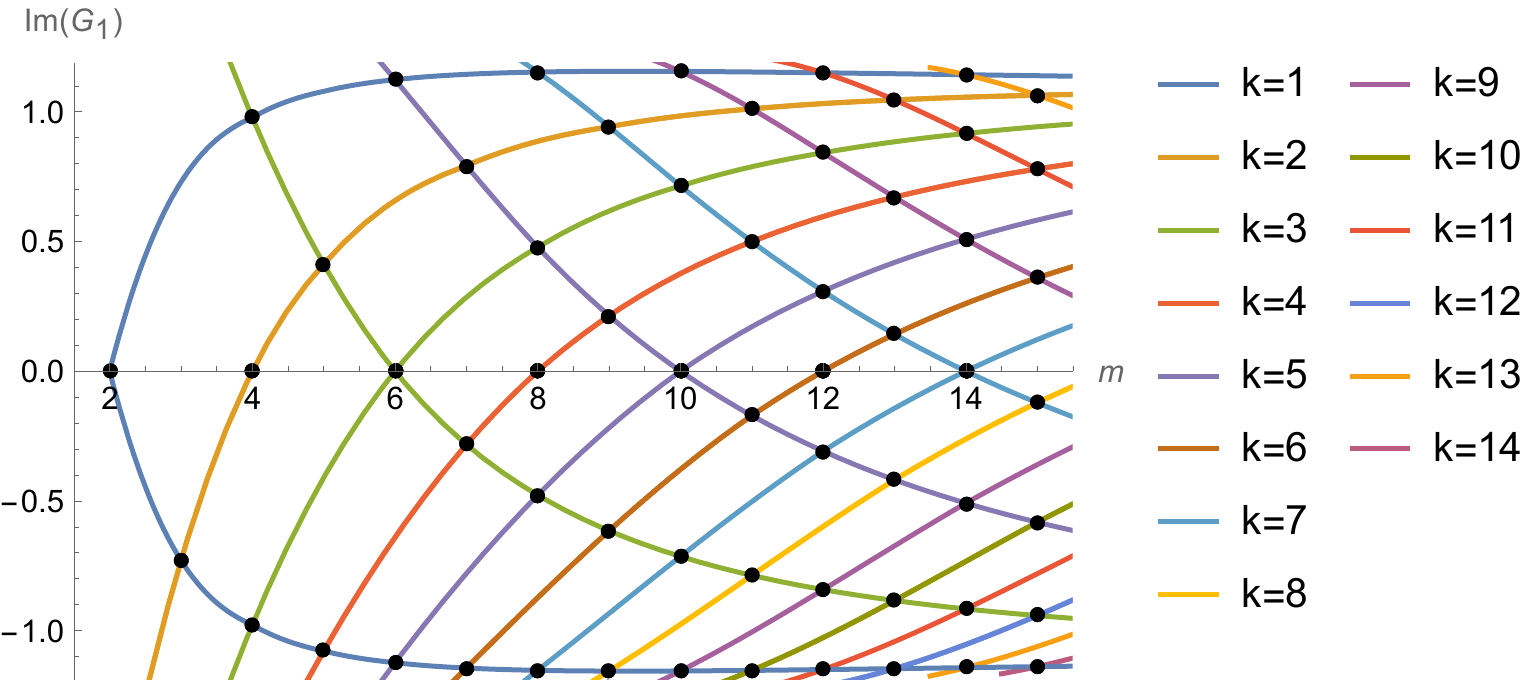}
	\caption{The purely imaginary solutions for $G_1$ in the $D=0$ non-Hermitian theories $\mathcal L=-(i^{m}/m)\, \phi^m$ from the general formulae \eqref{0D-general-sol}, \eqref{periodicity-general}, \eqref{non-degeneracy-general}. 
	Many of them are $\mathcal {PT}$ symmetric. 
	The exact solutions at integer $m$ are denoted by black dots.
	The colored curves from \eqref{0D-general-m} interpolate the integer-$m$ solutions with the same $k$, 
	according to $\a=e^{2\pi i\frac {k} m}$. 
}
	\label{0D-G1-plot}
\end{figure}

Let us discuss the more subtle situation of non-integer power $m$. 
We consider a family of non-Hermitian models with the Lagrangian $\mathcal L=-(i^{m}/m)\, \phi^m$, which is a $D=0$ version of the $\mathcal {PT}$-invariant theories proposed in \cite{Bender:1998ke}. 
In Fig. \ref{0D-G1-plot}, we present the purely imaginary solutions for $G_1$ from \eqref{0D-general-sol}, \eqref{periodicity-general}, \eqref{non-degeneracy-general}. 
The black dots indicate the integer-$m$ cases, 
where the periodicity conditions \eqref{periodicity-general} are clearly defined. 

We can also consider a rational power $m=(1/m_1)m_2$, 
where $(m_1$, $m_2)$ are positive and mutually prime integers. 
If we impose that the DS equations \eqref{0D-DS-general} are satisfied at generic $n$, 
then the principle of minimal singularity again leads to \eqref{0D-general-sol},  
but the periodicity conditions \eqref{periodicity-general} become   
$(\a^{1/m_1})^{m_2}=(\b^{1/m_1})^{m_2}=1$. 
For odd $m_1$, 
the imaginary $G_1$ solutions lie exactly on the interpolating curves for the integer-$m$ solutions. 
In Fig. \ref{0D-G1-plot},  
the interpolating curves are labelled by $k$ with $\a=e^{2\pi i\frac {k} m}$.  
Their general-$m$ analytic expression is
\be
\label{0D-general-m}
\text{Im}(G_1)=\pm \big|1+e^{2\pi i\frac {k}{m}}\big|\,m^{1/m}\left(\frac 1 m\right)_{1/m}\,,
\ee
where the signs of the even-$k$ cases depend on $m$. 
At large $m$, they asymptote to $\text{Im}(G_1)\rightarrow \pm 1$. 
For clarity, we do not plot the interpolating curves at $m<k$, 
as they oscillate more and more rapidly.  
In addition, many curves would intersect at the black dots 
due to $\a_{k}=\a_{k+m}$.

\section{One-dimensional theories} 
Are the $D>0$ exact solutions minimally singular? 
To address this question at $D=1$, 
we use the Hamiltonian formulation to reduce the number of free parameters. 
For an eigenstate with energy $E$, the expectation values satisfy some self-consistent equations 
\be
\label{1D-recursion}
\<\mathcal OH\>=E\<\mathcal O\>=\<H\mathcal O\>\,,
\ee
where the inner product is assumed to be compatible with the symmetry of the Hamiltonian. 
These self-consistency relations are the counterpart of the DS equations in the Lagrangian formulation.  
We will consider the basic examples of the Hermitian quartic and non-Hermitian cubic theories. 
\footnote{As in the $D=0$ case, 
we can also consider higher integer powers, as well as non-integer powers \cite{Li:2024rod}
. }

\subsection{Quartic theory}
We consider the massive quartic theory \cite{Bender:1969si}
\be
\label{1D-quartic}
H=p^2+\l x^2+g x^4\,,
\ee
which is related to the Ising universality class at higher $D$. 
We will show that the principle of minimal singularity manifests as the merging of different branches of $G_n$. 
The position operator $x$ and the momentum operator $p$ satisfy 
the canonical commutation relation $[x,p]=i\hbar$. 
Using the standard Hermitian inner product, 
one can formulate a self-consistent system in terms of the expectation values  \eqref{1D-recursion} 
and derive a recursion relation \cite{Han:2020bkb}:
\be
\label{1D-quartic-recursion}
&&\hbar^2\,(n+1)_3\, G_{n}+4E(n+3)\,G_{n+2}-4\l(n+4)\,G_{n+4}
\nn&=&4g(n+5)\,G_{n+6}\,,
\ee
where $G_n=G_n(t_1,t_2,\dots,t_n)|_{t_i\rightarrow t_1}=\<x^n\>$ 
is the equal-time limit of the $n$-point Green's function. 
\footnote{In higher spacetime dimensions, 
the $n$ analytic continuation can be performed in the coincidence limit or the equal-space-time limit of the Green's functions, where all the points have the same space-time coordinates. In other words, the 1-point functions of the composite operators $\phi^n$ are viewed as an analytic function in $n$. 
}
The normalization is set by $G_0=1$. 
Assuming that the solutions are parity invariant, 
we can focus on the even-$n$ Green's functions, as the odd-$n$ cases vanish. 
Instead of imposing positivity constraints \cite{Han:2020bkb}, 
we will study \eqref{1D-quartic-recursion} by analytic continuation in $n$ \cite{Li:2022prn}. 

For positive integer $n$, 
we can express $G_{n}$ in terms of $E$ and $G_2$ using \eqref{1D-quartic-recursion}. 
In Fig. \ref{3-branches}, we consider some cases of $(E, G_2)$ around the exact values, 
where $G_n$ has been divided by the leading asymptotic 
behavior $3^{n/3}n^{1/6}  \left[\G\left(\frac{n}{6}\right)\right]^2$. 
At large $n$, the solutions for $G_n$ exhibit three oscillatory branches,  
corresponding to $G_{6n'+2k}$ with $k=0,1,2$. 
As $(E, G_2)$ approach the exact values,  
the three branches of $G_n$ merge into one smooth curve in a considerable range! 

In fact, we have seen the merging phenomenon at $D=0$.  
In \eqref{D0-quartic-exact-sol}, 
the prefactors of $G_{4p}$ and $G_{4p+2}$ coincide for 
the $(+1)$ solution, so the two branches of solutions merge into one. 
For the quartic one-matrix model, 
the Green's functions 
also merge into one smooth curve 
at the one-cut solutions \cite{Brezin:1977sv}, 
which is discussed in more detail in Appendix \ref{Appendix_matrix-model} .  

\begin{widetext}
\onecolumngrid
\begin{figure}[h]
	\centering
	\begin{subfigure}{0.3\textwidth}
		\raggedright
		\includegraphics[width=1\linewidth]{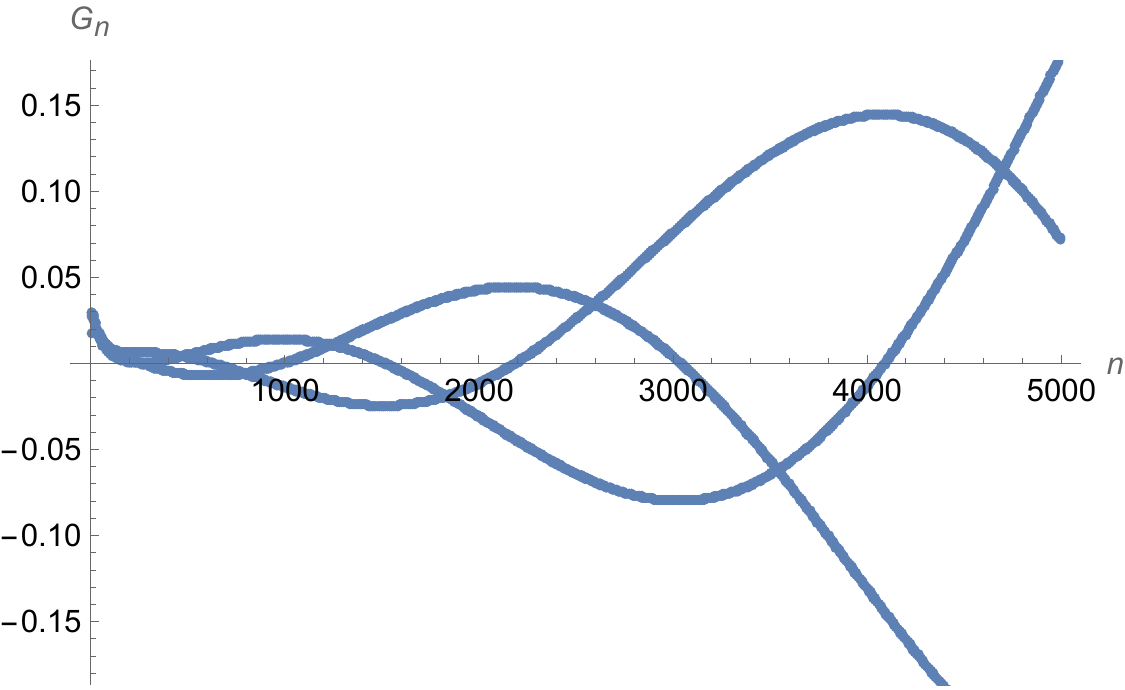}
		\caption{$(1.4, 0.3)$}
	\end{subfigure}
	\begin{subfigure}{.3\textwidth}
		\raggedright
		\includegraphics[width=1\linewidth]{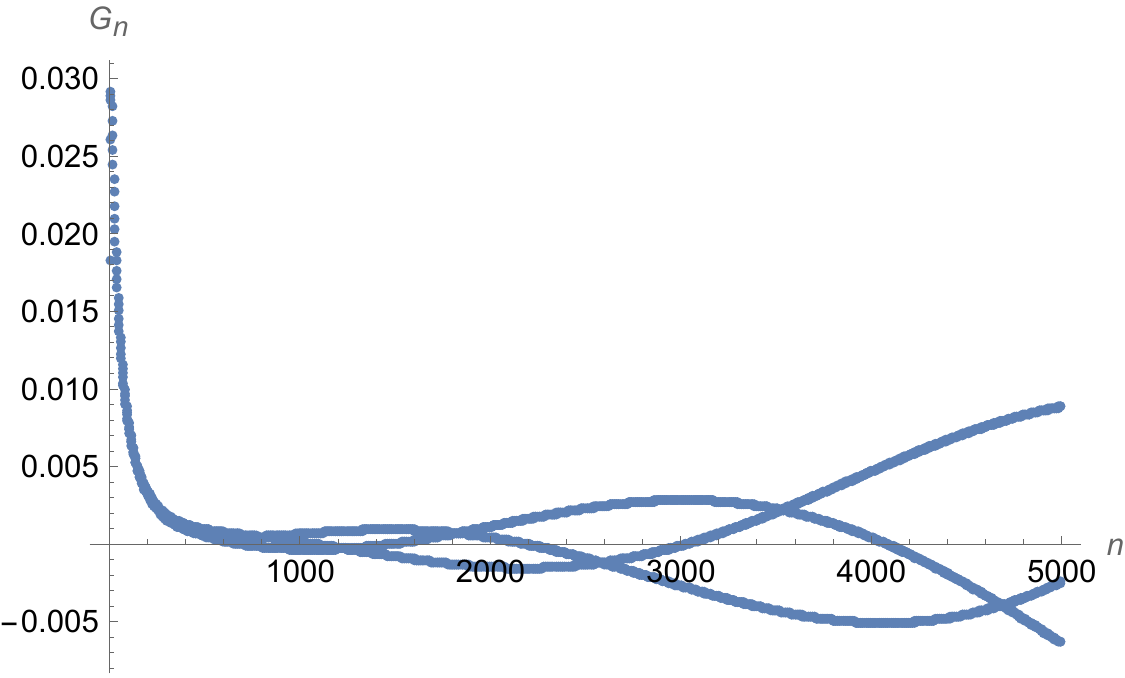}
		\caption{$(1.392, 0.306)$}
	\end{subfigure}
		\begin{subfigure}{.3\textwidth}
		\raggedright
		\includegraphics[width=1\linewidth]{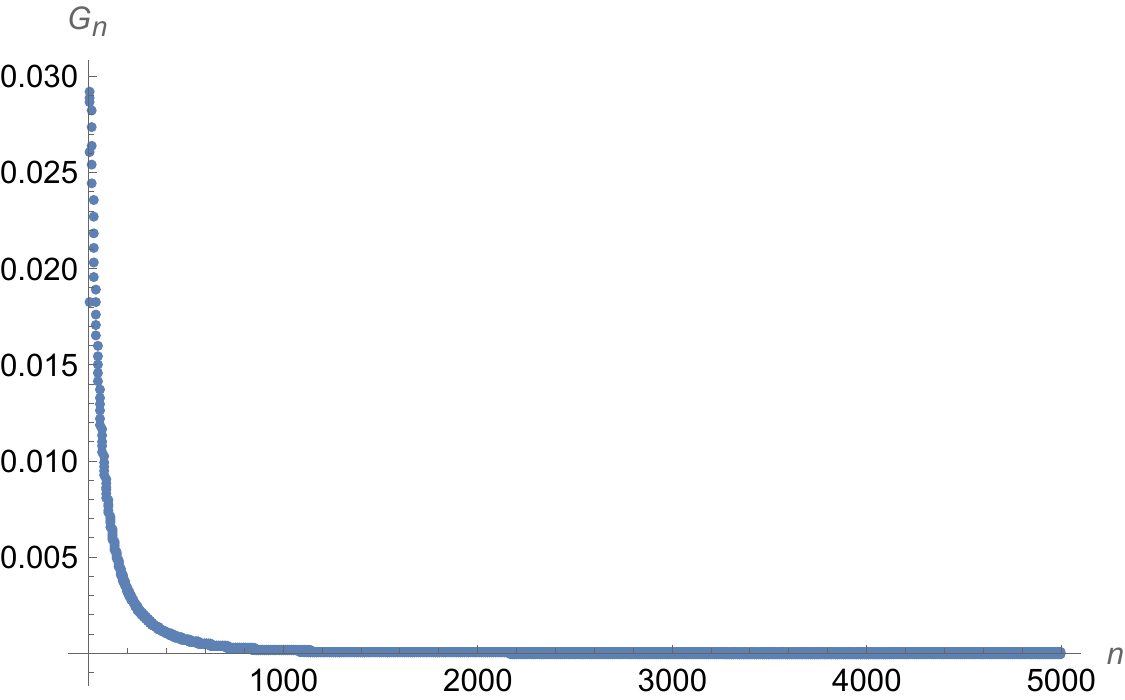}
		\caption{$(1.39235,0.30581)$}
	\end{subfigure}
	\begin{subfigure}{.3\textwidth}
		\raggedright
		\includegraphics[width=1\linewidth]{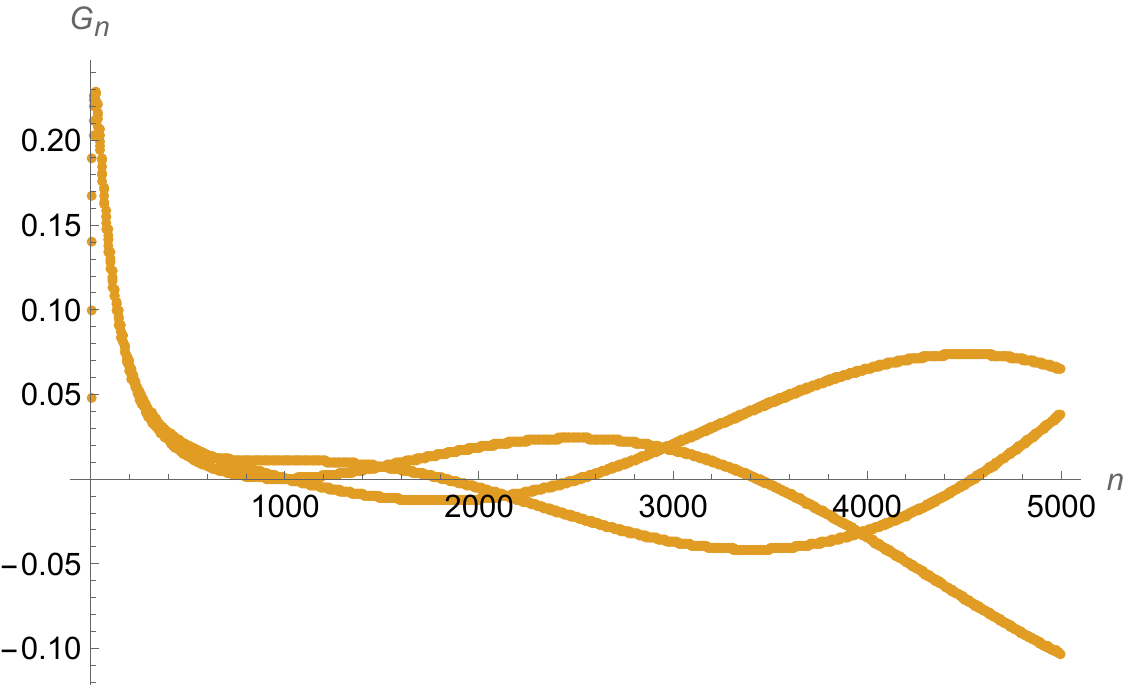}
		\caption{$(4.6, 0.8)$}
	\end{subfigure}
	\begin{subfigure}{.3\textwidth}
		\raggedright
		\includegraphics[width=1\linewidth]{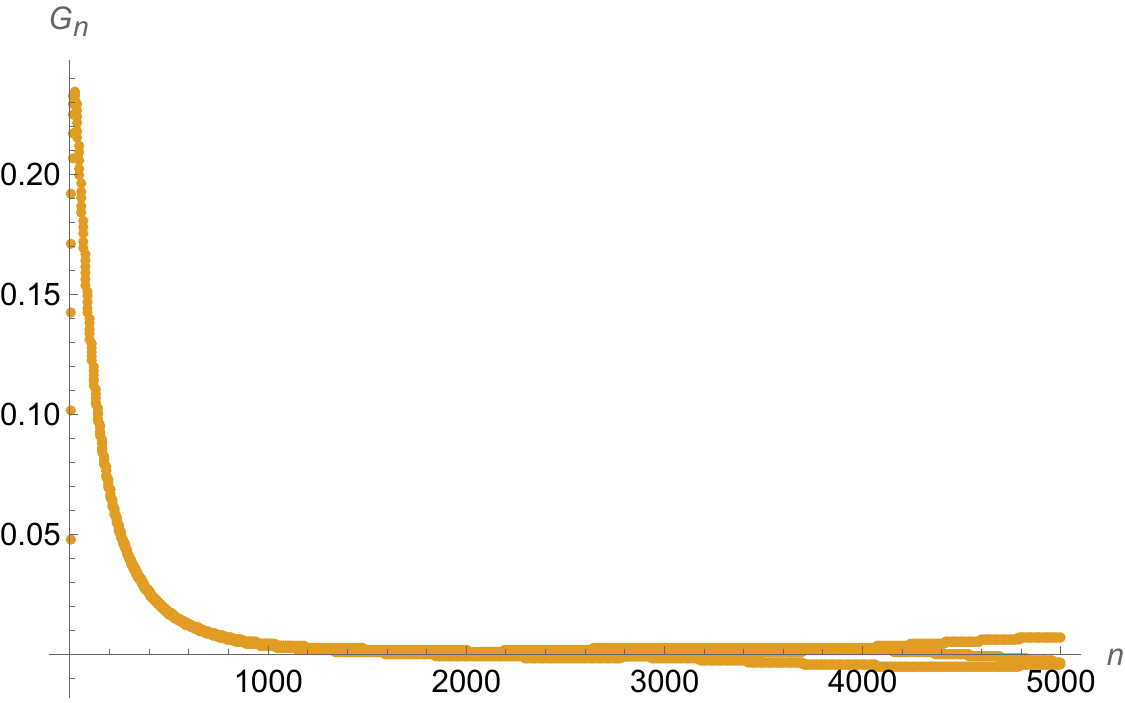}
		\caption{$(4.649, 0.801)$}
	\end{subfigure}
		\begin{subfigure}{.3\textwidth}
		\raggedright
		\includegraphics[width=1\linewidth]{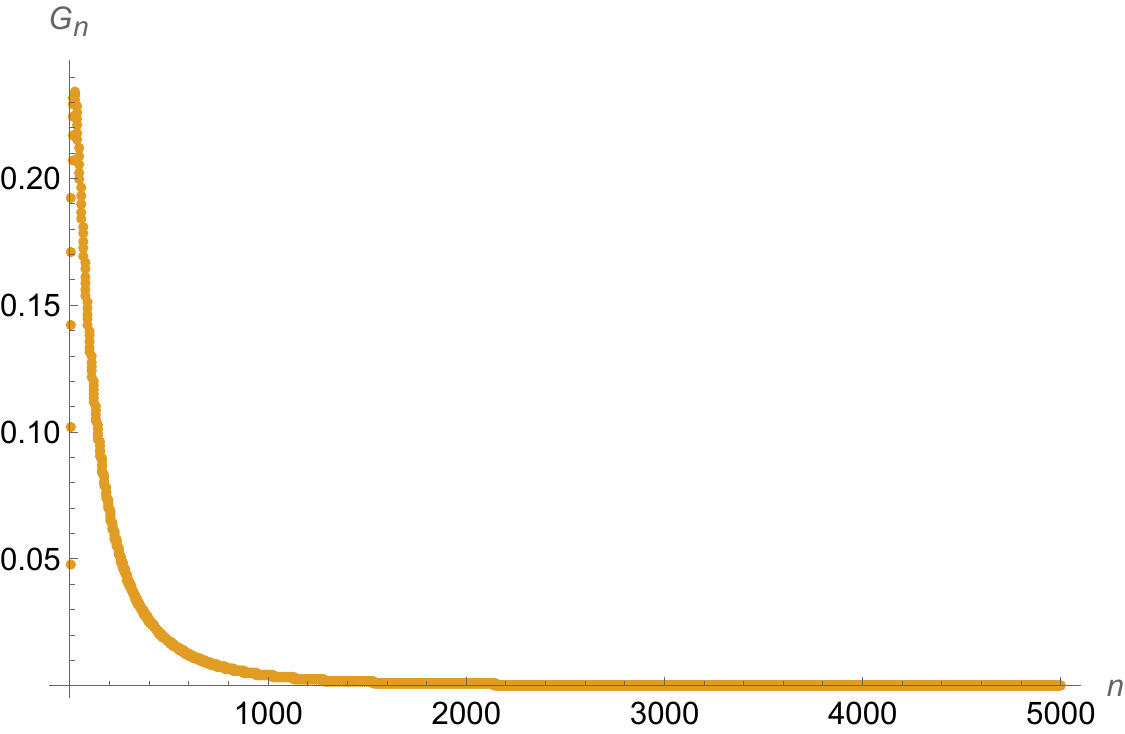}
		\caption{$(4.64881, 0.80125)$}
	\end{subfigure}
	\caption{The solutions of $G_n$ for the $D=1$ quartic theory \eqref{1D-quartic} from  
	\eqref{1D-quartic-recursion} with $\hbar=\l=g=1$. 
	The subfigures are labelled by the input $(E, G_2)$. 
	Around the exact values, the three branches of solutions merge into one decay curve at relatively small $n$.  
	At the same precision, the merging of the excited-state results (orange) extends to larger $n$ than the ground-state case (blue). }
	\label{3-branches}
\end{figure}
\end{widetext}

To show the existence of three branches of Green's functions, 
let us carry out the large-$n$ asymptotic analysis.  
If the Green's functions grow faster than $n$, 
the leading behavior is determined by 
\be
\label{1D-leading-recursion}
\hbar^2\,(n+1)_3\, G_{n}\sim 4g(n+5)\,G_{n+6}\quad (n\rightarrow \infty)\,,
\ee
where the second and third terms in \eqref{1D-quartic-recursion} have been omitted. 
This is a third-order difference equation for the even-$n$ Green's functions. 
At large $n$, the leading behavior of the even-$n$ Green's functions reads
\be
\label{1D-quartic-G-large-n}
G_n&\sim&
\left(\frac {9\hbar^2}{g}\right)^{\frac n 6} n^{1/6}\left[\G\left(\frac n 6\right)\right]^2
\sum_{k=0}^{2}\,c_k\,e^{2\pi i\frac {kn}{3}}
\quad (n\rightarrow \infty)\,.
\nn
\ee 
If \eqref{1D-leading-recursion} is solved exactly, 
we obtain the leading terms of the strong-coupling expansion in $1/g$. 
Below we set 
$\l=g=1$. 
The presence of the $c_1, c_2$ terms 
leads to three branches of $G_n$ at even $n$.  
In the merging limit, we have $c_1,c_2\rightarrow 0$ and 
two types of singular behaviors are removed, 
so the exact solutions are minimally singular.   
Then the third term in \eqref{1D-quartic-recursion} implies a subleading factor $\exp[-(n/2)^{1/3}]$, 
which is consistent with the decay behavior in Fig. \ref{3-branches}. 
Using \eqref{1D-recursion}, we can also deduce 
\be
\langle x^{n_1} p^{n_2}\rangle \sim i^{n_2}g^{{n_2}/6}{\left(\frac{n_1}{2}\right)}^{\frac{2{n_2}}{3}}G_{n_1}
\quad (n_1\rightarrow \infty)\,.\qquad
\ee 

The merging phenomenon in Fig. \ref{3-branches} indicates that 
the properties of the low-lying states are strongly constrained by the principle of minimal singularity. 
This is also consistent with the $E$ dependence in \eqref{1D-quartic-recursion}, 
which implies that the large-$n$ expansion is more accurate at smaller $E$. 
Let us introduce $F_n=G_{n+2}/G_n$.   
According to our new principle, 
$F_n$ should be minimally singular in $n$.  
If $F_n$ grows with $n$, then the leading behavior of $F_n$ is encoded in 
\be
\label{Fn-large-n}
F_{n}\,F_{n+2}\,F_{n+4}\sim\,g^{-1}\frac {\hbar^2\, n^2}{4}\quad (n\rightarrow \infty)\,,
\ee
which is similar to the $D=0$ cubic case \eqref{0D-DS-cubic-f}. 
Note that the large $n$ expansion is different from the small $\hbar$ expansion in the semiclassical WKB method. 
Below we further set 
$
\hbar=1$.  
The ``cube root'' of \eqref{Fn-large-n} gives three minimal solutions 
and two of them are complex. 
Only the real solution is consistent with the Hermitian inner product, 
which corresponds to \eqref{1D-quartic-G-large-n} with $c_1=c_2=0$.  
Then the systematic large-$n$ expansion gives
\be
\label{D1-quartic-Fn}
F_{n}&=&y^2
-\frac{1}{3}
-\frac 1 {2y^1}
+\frac{1+3E}{9y^2}
+\frac {7} {18y^3}
+\dots\Big|_{y=\sqrt[3]{\frac n 2}}\,. \qquad
\ee
It is not clear to us how to express \eqref{D1-quartic-Fn} in terms of  known  functions. 
Nevertheless, as the counterpart of the $(+1)$ solution in \eqref{D0-quartic-exact-sol}, 
we expect that the main implications of minimal singularity have been captured. 
Although nonperturbative corrections are not considered,  
a high-order truncation of the $1/n$ series 
\eqref{D1-quartic-Fn} 
can provide an accurate approximation for $F_n$
at sufficiently large $n$.

\begin{figure}[h]
	\centering
		\includegraphics[width=1\linewidth]{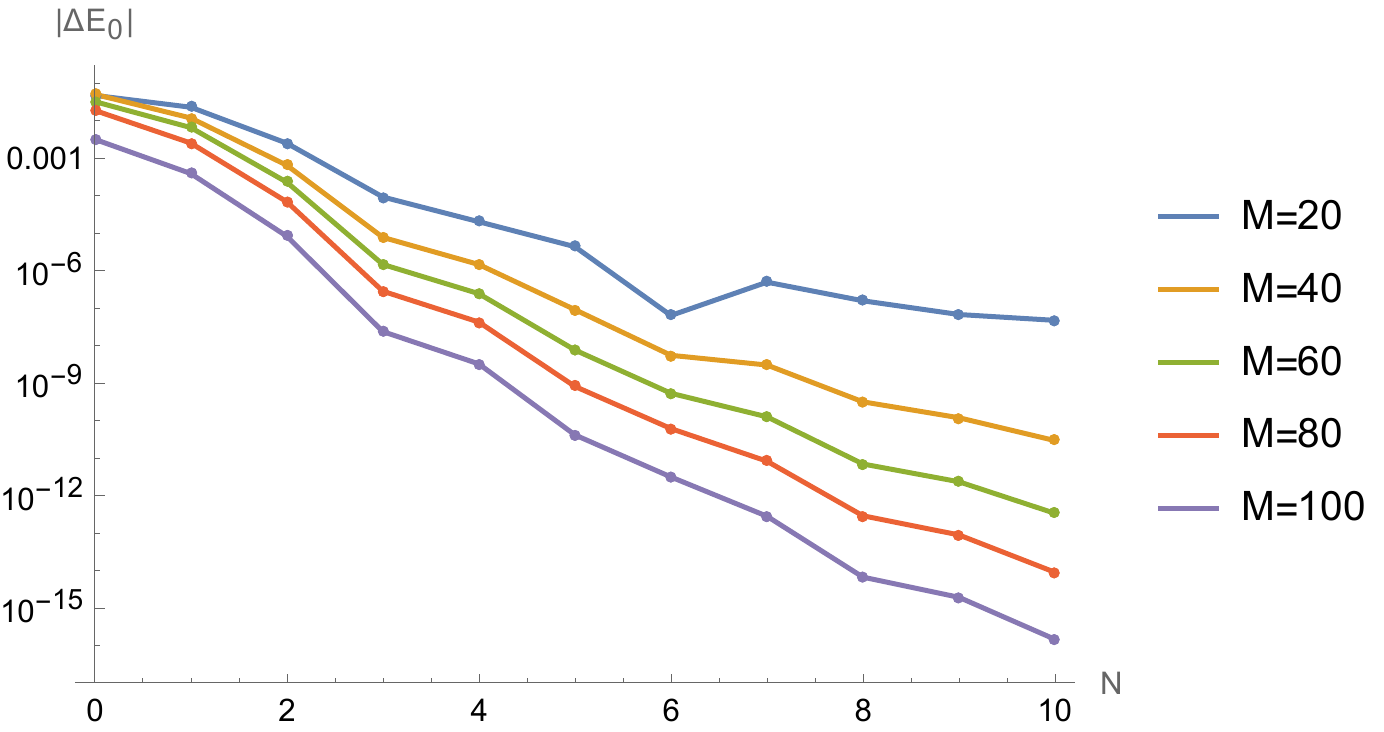}
	\caption{The absolute error in the ground-state energy $E_0$ of the $D=1$ quartic theory \eqref{1D-quartic} from the matching conditions \eqref{matching}, 
	where $\l=g=1$. Note that $M$ is the matching order. The $1/n$ series of $n^{-2/3}F_n$ is truncated to order $n^{-N}$. 
	The estimates converge rapidly to the exact value as $M, N$ increase. }
	\label{E0-error-plot}
\end{figure}

To determine $E$ and $G_2$, 
we will impose some matching conditions on $G_n$ at relatively large $n$. 
Above, the  $n^{-1}$ series \eqref{D1-quartic-Fn} encodes the constraints from the principle of minimal singularity. 
To connect with the observables in the nonperturbative regime with small $n$, 
we solve the recursion relation \eqref{1D-quartic-recursion} exactly to relatively large $n$.    
The analytic solutions for $G_n$ are given by high-degree polynomials in $E$ and $G_2$. 
Then we impose two matching conditions: 
\be
\label{matching}
G_{M+2}=F_{M}^{(N)}\,G_{M}\,,
\quad
G_{M+4}=F_{M+2}^{(N)}\,G_{M+2}\,,
\ee
where $M$ denotes the matching order and $N$ indicates the truncation order of the $1/n$ series. 
The real roots of the two polynomial equations provide accurate approximations for the low-lying observables. 
Fig. \ref{E0-error-plot} shows that the estimates of the ground-state energy $E_0$ improve rapidly with 
the matching order $M$ and the truncation order $N$.

Since the results for the higher states merge into one curve more easily,  
the principle of minimal singularity leads to more accurate results for the low-lying states. 
For example, the energy spectrum from the matching conditions \eqref{matching} with $(M,N)=(100,10)$ is $(1.39235164153029206,\,
4.6488127042152, \,
8.6550499661, \,$ $
13.15680465, \,
18.05715, \,
22.76, \,
28.69, \, 
39.2,\,\dots)$, 
where the last two digits of the approximate energies deviate from the exact values. 
The results for $G_2$ exhibit similar features. 

\subsection{Cubic theory}
Let us also revisit the $i\phi^3$ theory at $D=1$. 
In this case, the principle of minimal singularity is less explicit  
due to the absence of the merging phenomenon.  
The exact $\mathcal {PT}$ symmetric solutions exhibit five branches of $G_n$  
as in a generic self-consistent solution. 
However, we notice that  
they have only two types of singular behaviors at infinity, 
instead of all the five types. 
Therefore, the $\mathcal {PT}$ symmetric solutions 
can be extracted from the principle of minimal singularity as well.

The Hamiltonian of the $D=1$ non-Hermitian $\mathcal {PT}$-invariant cubic theory is
\cite{Bender:1998ke}
\be
H=p^2+i x^3\,.
\ee
We assume that the $\mathcal {PT}$ symmetry is unbroken and consider the $\mathcal {PT}$ inner product 
\be
\langle\psi_1|\psi_2\rangle^{\mathcal P\mathcal T}=C\int \mathrm{d}x\, [\psi_1(-x)]^\ast\psi_2(x)\,,
\ee
where $C$ is a normalization constant. 
We again have the self-consistency equations \eqref{1D-recursion} \cite{Li:2022prn},  
which give rise to a recursion relation for $G_n=\<x^n\>$:
\be
\label{1D-cubic-recursion}
(n+1)_3\,G_n+4E(n+3)\,G_{n+2}=2i(2n+9)\,G_{n+5}\,.\quad
\ee
At large $n$, the leading behavior is determined by 
\be
(n+1)_3\,G_n\sim 2i(2n+9)\,G_{n+5}\quad (n\rightarrow \infty)\,.
\ee
As a quintic-order difference equation, the leading asymptotic behavior of the general solution reads
\be
G_n&\sim&(-i)^n\Big(\frac 5 2\Big)^{\frac{2n}{5}}\Big[\G\big(\frac n 5\big)\Big]^2
\Big(c_1+c_2\cos\frac{2n\pi}{5}+c_3\sin\frac{2n\pi}{5}
\nn&&+c_4\cos\frac {4n\pi}{5}+c_5\sin\frac{4n\pi}{5}\Big)
\quad (n\rightarrow \infty)\,,
\ee
which has five types of singular behaviors at $n=\infty$. 
\footnote{Here we express the leading terms in terms of 
the trigonometric functions for notational simplicity. 
There are also five types of singular behaviors if we use the exponential functions. }

For $\mathcal {PT}$ symmetric solutions, $(c_1,\,c_2,\,\dots, \, c_5)$ should be real numbers. 
There are several minimally singular ansatz with at most two types of singular behaviors. 
It turns out that the solutions with real energies and a bounded-from-below spectrum are associated with $c_0=c_4=c_5=0$. 
The corresponding large $n$ expansion of $G_n$ is
\be
\label{1D-cubic-large-n-expansion}
G_n&=&(-i)^n\, n^{3/{10}}\left(\frac 5 2\right)^{\frac{2n}5}
\Big[\G\Big(\frac n 5\Big)\Big]^2
\Bigg\{(c_2+i c_3)\frac{e^{-2\pi i\frac{n}{5}}}{2}
\nn&&\times
\Big[1-2E\Big(\frac{-2}{n}\Big)^{\frac 1 5}+2E^2\Big(\frac{-2}{n}\Big)^{\frac 2 5}
+\dots\Big]+\text{c.c.}
\Bigg\}\,,\nn
\ee
where c.c. stands for complex conjugate and we have added a subleading factor $n^{3/{10}}$. 
To determine the free parameters $(E, G_1, c_2, c_3)$, we impose four matching conditions
\be
G_n^{(\text{n.p.})}=G_n^{(\text{p.})}\,,
\ee
where $n=M,\,M+1,\,M+2,\,M+3$. 
The nonperturbative expressions of $G_n^{(\text{n.p.})}$ are derived from the exact recursion relation \eqref{1D-cubic-recursion} at $n=-3,-2,\dots,M-2$. 
They are high-degree polynomials in $E$, but at most linear in $G_1$.  
The perturbative expressions of $G_n^{(\text{p.})}$ are obtained from the truncation of the large $n$ series to order $n^{-N}$. 
The first three leading terms are written explicitly in \eqref{1D-cubic-large-n-expansion}. 

As in the quartic case, the results converge rapidly to the exact values with the matching order $M$ and the truncation order $N$. 
For $(M,N)=(100,10)$, the real-energy solutions correspond to the $\mathcal {PT}$-symmetric ground state and first-excited state. 
The explicit results for $(E,G_1)$ are
$(1.1562670719881112, -0.59007253309070011 i)$ and $(4.10942, -0.982086 i)$, 
where the last two digits deviate from the exact values. 
In comparison to the quartic case, we obtain less energy levels because some solutions in the non-Hermitian cubic case are spurious and unstable. 
 
Although there are only two types of singular behaviors at $n=\infty$, the Green's functions still exhibit five branches of solutions. 
In fact, if we take into account the odd $n$ Green's functions in the Hermitian quartic case, 
the exact solutions also have two types of singular behaviors at infinity, 
as here in the cubic case. 

In the end, it seems that $(c_2, c_3)$ are not independent 
because $c_3=c_2\tan(-3\p/10)$ is satisfied to high precision near the exact solutions.

\section{Discussion}
In this work, we considered the analytic continuations of the Green's functions $G_n$ to complex $n$ and 
discovered the principle of minimal singularity. 
Besides good analytic properties, 
the minimality of a self-consistent solution is closely related to 
the simplicity of an asymptotic behavior at large $n$ \cite{Bender:2022eze,Bender:2023ttu}
and the irreducibility of an operator-algebra representation \cite{Li:2022prn}. 

We explained how to resolve the indeterminacy of self-consistency equations using this novel principle.  
At $D=0$, we obtained the closed-form solutions of the general $g\phi^m$ theory in 
\eqref{0D-general-sol}, \eqref{periodicity-general}, \eqref{non-degeneracy-general}.  
The minimal-singularity approach extends to $D=1$, where rapidly convergent results were obtained. 
At higher $D$, it would be interesting to explore the analytic continuations in 
both the number of fundamental fields and the number of derivatives. 
\footnote{In this work, we avoid the explicit time derivatives in the $D=1$ cases by considering the Hamiltonian formulation.}
It is also important to go beyond the coincidence limit. 
For instance, we can study the $n$ analytic continuation of $\<T\{\phi^n(y_1)\phi^n(y_2)\}\>$ or $\<T\{\phi(y_1)\phi(y_2)\phi^n(y_3)\}\>$. 
See the recent work \cite{Guo:2023qtt} for some perturbative examples in the context of conformal field theory.  
In analogy with the connection between the singularity structure in angular momentum and 
the asymptotic behavior of scattering amplitudes, 
it would be interesting to examine the connection between the singularity structure in $n$ and the analytic properties of Green's functions in coordinate or in momentum space. 
\footnote{The principle of minimal singularity may also apply to other parameters, such as coordinate space and momentum space. 
For example, the operator decoupling phenomena in the Ising conformal field theory 
\cite{El-Showk:2012cjh,El-Showk:2014dwa} can be interpreted as 
a minimization of the complexity of the lightcone singularity structure in coordinate space. }

In the recursion relation for consecutive Green's functions, 
the exact solutions are related to minimally singular coefficient functions $F_n$. 
The null state approach \cite{Li:2023nip, Li:2022prn} can be viewed as 
a rational approximation for the exact recursion relation. 
Without a priori knowledge of $F_n$, 
a truncated null state condition leads to an approximate recursion relation for multiple Green's functions, 
where the coefficients are constant  and determined unbiasedly by the self-consistent equations.  
Using the null state condition, one can also reconstruct the unequal-time Green's functions  
from the equal-time limit 
 \cite{Li:2023nip,Guo:2023gfi}. 

If we consider the DS equations at $D>0$, there will be infinitely many free parameters. 
For the emergence of a proper inner product, they need to exhibit better analytic behavior than an arbitrary set of numbers.  
The null state approach is a particularly useful way to derive the minimally singular solutions. 
To further extract physically more meaningful solutions, we can impose additional constraints, 
such as positive semidefiniteness from unitarity \cite{Anderson:2016rcw,Lin:2020mme,Kazakov:2021lel,Kazakov:2022xuh} 
or spectral boundedness from stability \cite{Li:2023nip}. 
\footnote{It would be interesting to develop an approach that is directly based on the spectral boundedness. 
Although a stronger assumption will rule out other potentially useful solutions, 
the corresponding method can be more efficient if the size of the solution space is significantly reduced. }

In gauge theories, 
it may also be crucial to study the Green's functions without relying on positivity constraints. 
In quantum chromodynamics, the color singlet Green's functions are expected to be positive semidefinite, but the non-color-singlet Green's functions can violate positivity constraints. 
This is closely related to the perspective on the confinement mechanism 
in terms of the absence of a mass pole on the real, positive $p^2$-axis for non-color-singlet Green's functions. 
\footnote{It could be subtle to extract gauge-invariant information from a gauge-dependent Green's function. }
(See e.g. \cite{Cornwall:1980zw,Munczek:1983dx,Krein:1990sf} and references therein.)
For fermionic degrees of freedom, 
we may consider the analytic continuation in the power of the bilinear operators, such as the simplest case $(\bar\psi\psi)^n$. 
For $n=1$, the chiral condensate $\<\bar\psi\psi\>$ plays a special role in the spontaneous breaking of the chiral symmetry. 
As suggested in \cite{Li:2017agi,Li:2017ukc}, it would also be interesting to bootstrap the Shifman-Vainshtein-Zakharov sum rules \cite{Shifman:1978bx,Shifman:1978by}. 
It is certainly also important to go beyond the vacuum condensates. 
For example, we can consider $\<T\{[\bar\psi(\bar\psi\psi)^n](y_1)\,[\psi(\bar\psi\psi)^n](y_2)\}\>$, which is the $n$ generalization of the 2-point quark-antiquark Green's function. 

We believe that the principle of minimal singularity has  
broad applicability to self-consistent problems. 
The fact that the low-lying properties are severely constrained by this principle   
resonates with the remarkable progress in the conformal bootstrap program \cite{Poland:2018epd}. 
\footnote{In light of the results in this work, 
we revisit the conformal bootstrap approach to the 3D Ising model in \cite{Li:2023tic}. }
Our novel principle also sheds light on the non-positive bootstrap studies of strongly coupled systems.
Their developments are much slower 
as the powerful positivity constraints are not applicable. 
The principle of minimal singularity should be one of the missing pieces.  

\section*{Acknowledgments}
I would like to thank the referee for the insightful and constructive comments and suggestions. 
This work was supported by the 100 Talents Program of Sun Yat-sen University, 
the Natural Science Foundation of China (Grant No. 12205386) and the Guangzhou Municipal
Science and Technology Project (Grant No. 2023A04J0006).

\appendix

\section{Merging phenomenon in the quartic matrix model}
\label{Appendix_matrix-model}
In this appendix, we present the details of the merging phenomenon in the quartic matrix model. 
Here the exact one-cut solutions are precisely the minimally singular solutions. 
We consider the Hermitian one-matrix model with the potential
\be
V(x)=\frac 1 2 x^2+\frac g 4 x^4\,.
\ee 
In the large $N$ limit, the main Green's functions are the single-trace moments
\be
G_n=\langle \text{Tr}M^n\rangle=\frac {1}{N}\frac {\int {dM}\,\text{tr}M^n\,e^{-N\text{tr}V(M)}}
{\int {dM}\,e^{-N\text{tr}V(M)}},
\ee
where $M$ is an $N\times N$ Hermitian matrix. 
We have used the normalized trace so that $G_0=\text{Tr}I=1$. 
In the large $N$ limit, the Dyson-Schwinger equations are
\be
G_{n+1}+g\,G_{n+3}=\sum_{p=0}^{n-1}\,G_p\, G_{n-p-1}\,,
\label{DS-matrix}
\ee
which are also known as the loop equations in the context of matrix models. 
The right hand side involves the products of two Green's functions due to the large $N$ factorization. 
Following the standard procedure, we introduce the resolvent
\be
R(z)=\left\langle\text{Tr}\, \frac{1}{z-M}\right\rangle=\sum_{n=0}^{\infty}G_n\,z^{-n-1}\,.
\ee
The DS equations \eqref{DS-matrix} imply a quadratic equation for the resolvent. 
The solution reads
\be
R(z)=\frac{V'(z)}{2}-\sqrt{\frac{V'(z)^2}{4}-(gz^2+gG_1z+gG_2+1)}\,,
\nn
\label{resolvent}
\ee
where the relevant solution is selected by the condition $\lim_{z\rightarrow \infty}[zR(z)]=1$. 
If we assume that the solutions are parity symmetric, 
then $G_1=0$. The only free parameter is $G_2$. 

The eigenvalue distribution is related to the singularity structure of $R(z)$, 
which has branch point singularities. 
They are associated with the roots of the polynomial in the square root in \eqref{resolvent}. 
In general, a degree-six polynomial has six different roots and thus there should be three branch cuts. 
However, the number of branch cuts can be reduced if some roots are at the same point. 
In Brezin-Itzykson-Parisi-Zuber's classical work \cite{Brezin:1977sv}, 
the one-cut solutions of the quartic one-matrix model were constructed. 
Here we consider a slightly more general form
\be
R^{(\text{one-cut})}(z)=\frac{V'(z)}{2}-a_0(z^2+a_1)\sqrt{z^2+a_2}\,,
\ee
where $a_0>0$, and $(a_0, a_1, a_2)$ are functions of $g$. 
The one-cut solutions for $G_2$ are given by
\be
G_2^{(\text{one-cut,$\pm$})}=\frac{(\pm1)(12g+1)^{3/2}-18g-1}{54g^2}\,.
\label{matrix-G2-one-cut}
\ee
They are precisely the minimally singular solutions. 
We will show that the Green's functions exhibit the merging phenomenon  
as the free parameter $G_2$ approaches the one-cut solutions. 
The $(+)$ one-cut solution of $G_2$ is positive for $g\geq -\frac {1}{12}$. 
The two one-cut solutions meet at the critical point $g=-\frac {1}{12}$ and 
become a complex conjugate pair for $g<-\frac {1}{12}$.  

\begin{widetext}
\onecolumngrid
\begin{figure}[h]
	\centering
		\begin{subfigure}{.3\textwidth}
		\raggedright
		\includegraphics[width=1\linewidth]{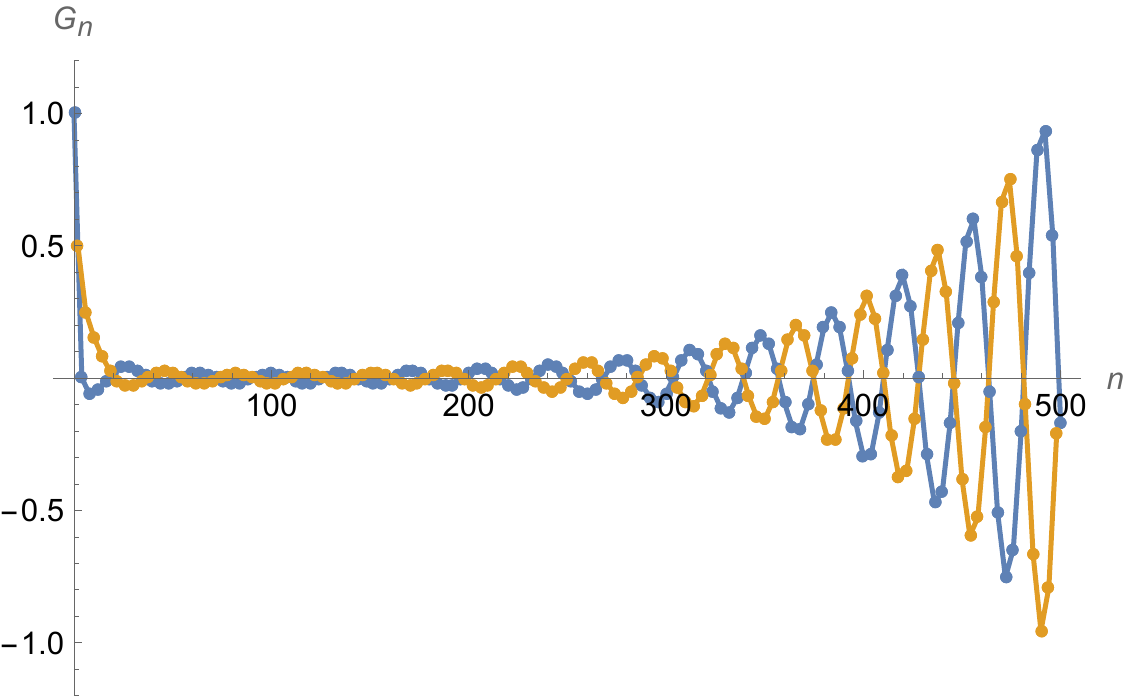}
		\caption{$G_2=1>G_2^\text{(one-cut,+)}$}
	\end{subfigure}
	\begin{subfigure}{.3\textwidth}
		\raggedright
		\includegraphics[width=1\linewidth]{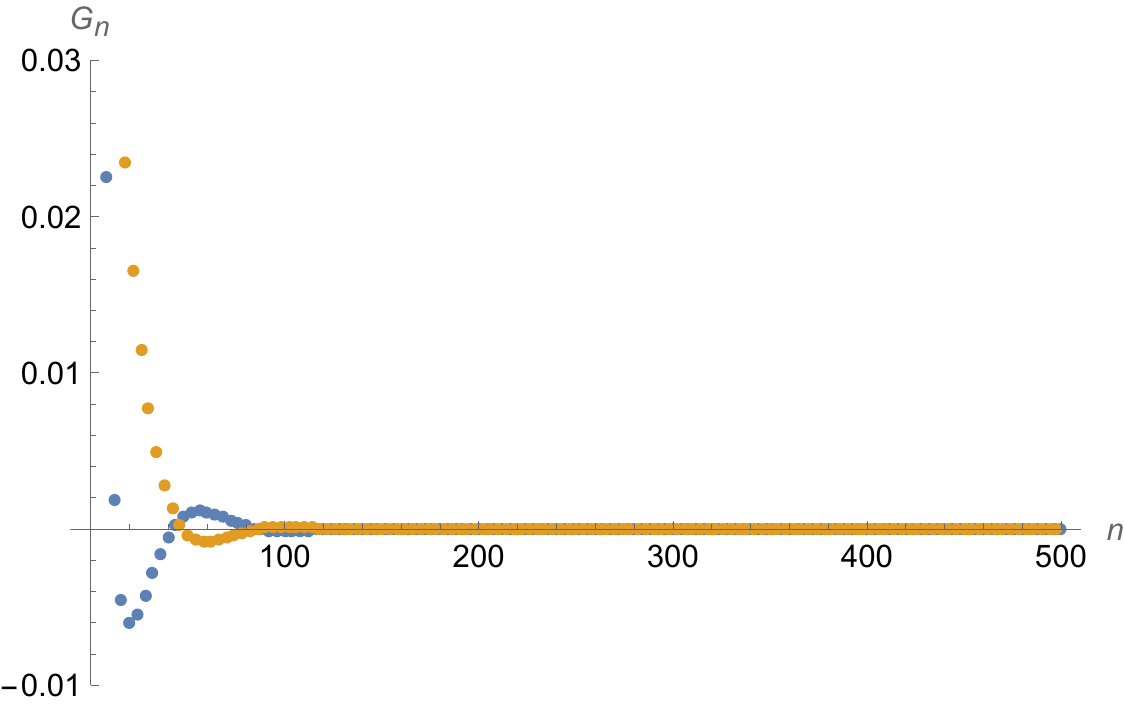}
		\caption{$G_2=0.6>G_2^\text{(one-cut,+)}$}
	\end{subfigure}
		\begin{subfigure}{.3\textwidth}
		\raggedright
		\includegraphics[width=1\linewidth]{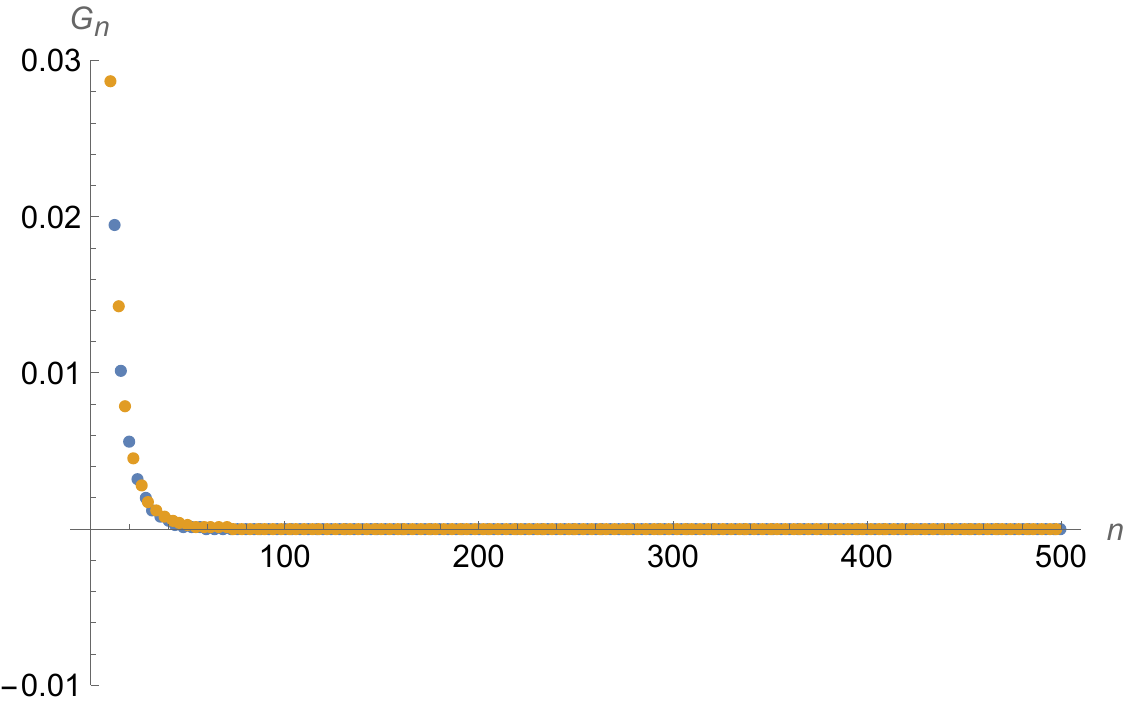}
		\caption{$G_2=0.517>G_2^\text{(one-cut,+)}$}
	\end{subfigure}
	\begin{subfigure}{0.3\textwidth}
		\raggedright
		\includegraphics[width=1\linewidth]{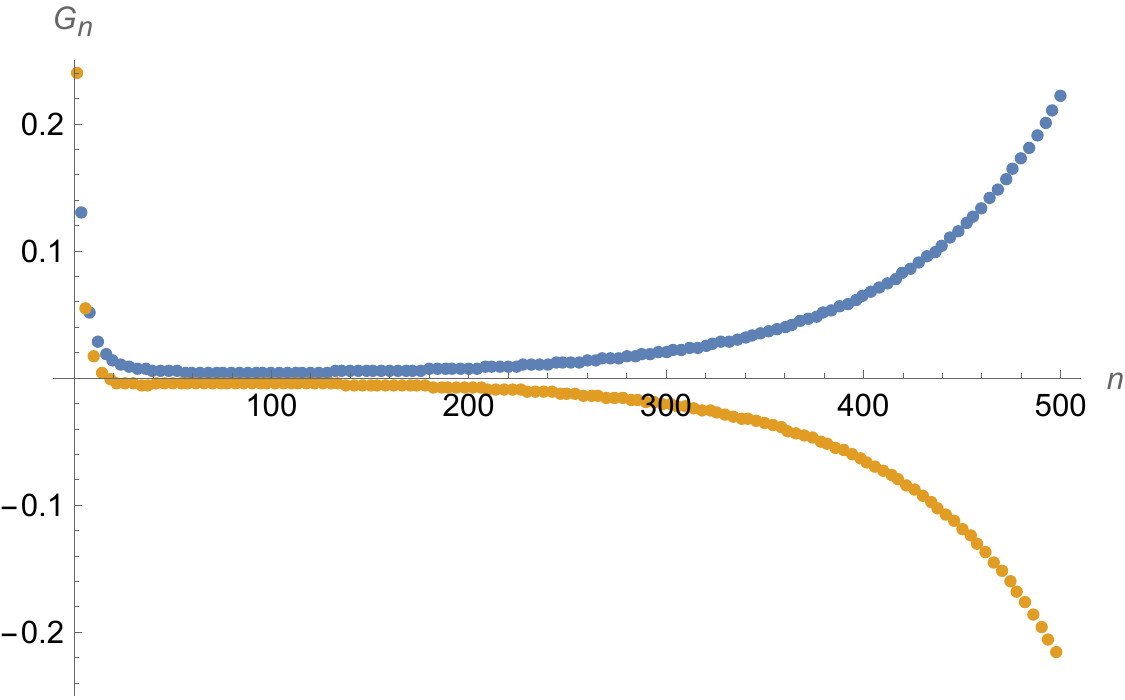}
		\caption{$G_2=0.48<G_2^\text{(one-cut,+)}$}
	\end{subfigure}
	\begin{subfigure}{.3\textwidth}
		\raggedright
		\includegraphics[width=1\linewidth]{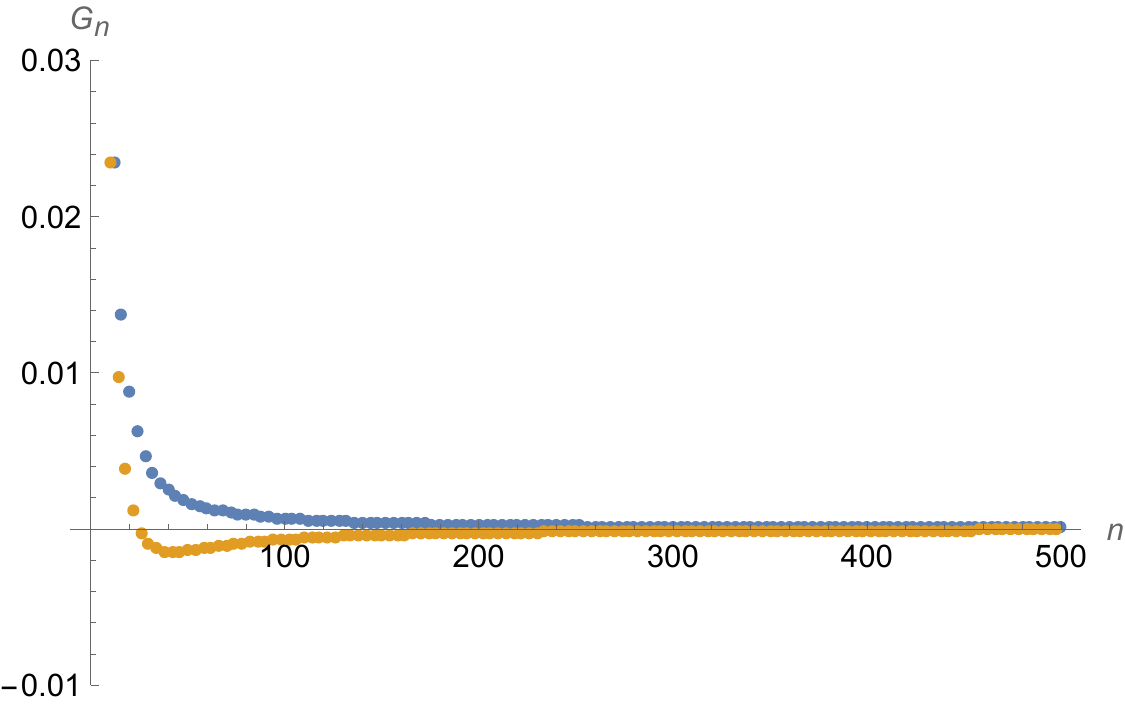}
		\caption{$G_2=0.5<G_2^\text{(one-cut,+)}$}
	\end{subfigure}
		\begin{subfigure}{.3\textwidth}
		\raggedright
		\includegraphics[width=1\linewidth]{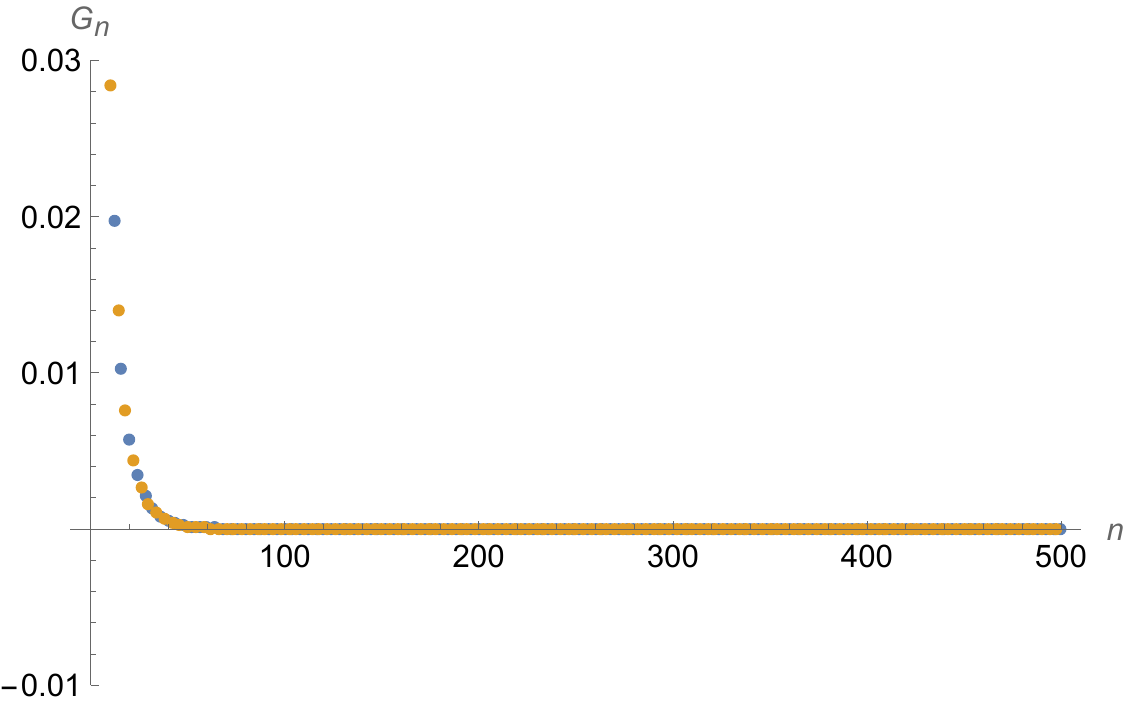}
		\caption{$G_2=0.516<G_2^\text{(one-cut,+)}$}
	\end{subfigure}

	\caption{The even-$n$  Green's functions for the $g=1$ quartic matrix model with different $G_2$. 
	As the input $G_2$ approaches the one-cut value $G_2^\text{(one-cut,+)}
	=0.51615...$ from above or below,  
	the two branches of Green's functions merge into one smooth curve.  
	We have divided $G_n$ by $2^{n/2}$. 
	Here $G_{4p}$ and $G_{4p+2}$ with $p=0,1,2,\dots$ are denoted by the blue and orange dots. 
	For $G_2=1$, the dots are joined so that the oscillatory behaviors are more clear. 
	}
	\label{fig-quartic-matrix}
\end{figure}

\end{widetext}

For $g>0$, 
the $(+)$ solution corresponds to the more standard case with $a_2<0$, 
where the branch cut is on the real axis and the eigenvalues have finite support at real values. 
In Fig. \ref{fig-quartic-matrix}, 
we consider the concrete case of $g=1$.  
One can see that the two branches of Green's functions merge into one smooth curve 
as the input parameter $G_2$ approaches the one-cut value 
$G_2^\text{(one-cut,+)}$ from above or below. 
When $G_2$ is greater than $G_2^\text{(one-cut,+)}$, the two branches of Green's functions exhibit oscillatory behaviors. 

We also examine the behaviors of Green's functions around the other one-cut solution, i.e., $G_2^\text{(one-cut,-)}$, 
which is negative at $g=1$.  
Since $a_2>0$, the eigenvalues in this case have finite support on the imaginary axis. 
Nevertheless, we find that $G_{4p}$ and $-G_{4p+2}$ merge into one curve in the one-cut limit, which is similar to the $D=0$ scalar theory solution in \eqref{D0-quartic-exact-sol}.  
We could also consider the one-cut solutions with $G_1\neq 0$. 
As in the case of the $\mathcal {PT}$-invariant scalar theories, there exist purely imaginary solutions for $G_1$ when $g<-\frac{1}{12}$. 

In the null state approach \cite{Li:2023nip}, we can also derive the one-cut and two-cut solutions of the one-matrix models, including the complex cases where the branch cuts are not on the real axis.  
The requirement is that the vanishing behaviors on the branch cuts 
with eigenvalue-distribution support admit polynomial approximations. 
\footnote{The vanishing behavior for the case with positive weight on the real axis was discussed previously in \cite{Lin:2020mme}. }
In this way, the null state condition approach can be viewed as a rational-approximation realization of the minimal singularity principle.


\begin{thebibliography}{10}

\bibitem{Yang:1952be}
C.~N.~Yang and T.~D.~Lee,
``Statistical theory of equations of state and phase transitions. 1. Theory of condensation,''
Phys. Rev. \textbf{87}, 404-409 (1952)
doi:10.1103/PhysRev.87.404

\bibitem{Lee:1952ig}
T.~D.~Lee and C.~N.~Yang,
``Statistical theory of equations of state and phase transitions. 2. Lattice gas and Ising model,''
Phys. Rev. \textbf{87}, 410-419 (1952)
doi:10.1103/PhysRev.87.410

\bibitem{Wick:1954eu}
G.~C.~Wick,
``Properties of Bethe-Salpeter Wave Functions,''
Phys. Rev. \textbf{96}, 1124-1134 (1954)
doi:10.1103/PhysRev.96.1124


\bibitem{Regge:1959mz}
T.~Regge,
``Introduction to complex orbital momenta,''
Nuovo Cim. \textbf{14}, 951 (1959)
doi:10.1007/BF02728177

\bibitem{Brout:1959zz}
R.~Brout,
``Statistical Mechanical Theory of a Random Ferromagnetic System,''
Phys. Rev. \textbf{115}, 824-835 (1959)
doi:10.1103/PhysRev.115.824

\bibitem{deGennes:1972zz}
P.~G.~de Gennes,
``Exponents for the excluded volume problem as derived by the Wilson method,''
Phys. Lett. A \textbf{38}, 339-340 (1972)
doi:10.1016/0375-9601(72)90149-1

\bibitem{Fortuin:1971dw}
C.~M.~Fortuin and P.~W.~Kasteleyn,
``On the Random cluster model. 1. Introduction and relation to other models,''
Physica \textbf{57}, 536-564 (1972)
doi:10.1016/0031-8914(72)90045-6

\bibitem{Bollini:1972bi}
C.~G.~Bollini and J.~J.~Giambiagi,
``Lowest order divergent graphs in nu-dimensional space,''
Phys. Lett. B \textbf{40}, 566-568 (1972)
doi:10.1016/0370-2693(72)90483-2

\bibitem{tHooft:1972tcz}
G.~'t Hooft and M.~J.~G.~Veltman,
``Regularization and Renormalization of Gauge Fields,''
Nucl. Phys. B \textbf{44}, 189-213 (1972)
doi:10.1016/0550-3213(72)90279-9


\bibitem{Wilson:1971dc}
K.~G.~Wilson and M.~E.~Fisher,
``Critical exponents in 3.99 dimensions,''
Phys. Rev. Lett. \textbf{28}, 240-243 (1972)
doi:10.1103/PhysRevLett.28.240

\bibitem{Dyson:1949ha}
F.~J.~Dyson,
``The S matrix in quantum electrodynamics,''
Phys. Rev. \textbf{75}, 1736-1755 (1949)
doi:10.1103/PhysRev.75.1736

\bibitem{Schwinger:1951ex}
J.~S.~Schwinger,
``On the Green's functions of quantized fields. 1.,''
Proc. Nat. Acad. Sci. \textbf{37}, 452-455 (1951)
doi:10.1073/pnas.37.7.452

\bibitem{Schwinger:1951hq}
J.~S.~Schwinger,
``On the Green's functions of quantized fields. 2.,''
Proc. Nat. Acad. Sci. \textbf{37}, 455-459 (1951)
doi:10.1073/pnas.37.7.455

\bibitem{Bender:1988bp}
C.~M.~Bender, F.~Cooper and L.~M.~Simmons,
``Nonunique Solution to the Schwinger-dyson Equations,''
Phys. Rev. D \textbf{39}, 2343-2349 (1989)
doi:10.1103/PhysRevD.39.2343

\bibitem{Bender:2022eze}
C.~M.~Bender, C.~Karapoulitidis and S.~P.~Klevansky,
``Underdetermined Dyson-Schwinger Equations,''
Phys. Rev. Lett. \textbf{130}, no.10, 101602 (2023)
doi:10.1103/PhysRevLett.130.101602
[arXiv:2211.13026 [math-ph]].

\bibitem{Bender:2023ttu}
C.~M.~Bender, C.~Karapoulitidis and S.~P.~Klevansky,
``Dyson-Schwinger equations in zero dimensions and polynomial approximations,''
[arXiv:2307.01008 [math-ph]].

\bibitem{Li:2023nip}
W.~Li,
``Taming Dyson-Schwinger Equations with Null States,''
Phys. Rev. Lett. \textbf{131}, no.3, 031603 (2023)
doi:10.1103/PhysRevLett.131.031603
[arXiv:2303.10978 [hep-th]].

\bibitem{Anderson:2016rcw}
P.~D.~Anderson and M.~Kruczenski,
``Loop Equations and bootstrap methods in the lattice,''
Nucl. Phys. B \textbf{921}, 702-726 (2017)
doi:10.1016/j.nuclphysb.2017.06.009
[arXiv:1612.08140 [hep-th]].

\bibitem{Lin:2020mme}
H.~W.~Lin,
``Bootstraps to strings: solving random matrix models with positivity,''
JHEP \textbf{06} (2020), 090
[arXiv:2002.08387 [hep-th]].

\bibitem{Kazakov:2021lel}
V.~Kazakov and Z.~Zheng,
``Analytic and numerical bootstrap for one-matrix model and \textquotedblleft{}unsolvable\textquotedblright{} two-matrix model,''
JHEP \textbf{06} (2022), 030
[arXiv:2108.04830 [hep-th]].

\bibitem{Kazakov:2022xuh}
V.~Kazakov and Z.~Zheng,
``Bootstrap for lattice Yang-Mills theory,''
Phys. Rev. D \textbf{107}, no.5, L051501 (2023)
doi:10.1103/PhysRevD.107.L051501
[arXiv:2203.11360 [hep-th]].

\bibitem{Rattazzi:2008pe}
R.~Rattazzi, V.~S.~Rychkov, E.~Tonni and A.~Vichi,
``Bounding scalar operator dimensions in 4D CFT,''
JHEP \textbf{12}, 031 (2008)
doi:10.1088/1126-6708/2008/12/031
[arXiv:0807.0004 [hep-th]].




\bibitem{Chew:book}
G.~F.~Chew, 
"The S-matrix theory of strong interactions," 
(W. A. Benjamin, Inc., New York, 1961).

\bibitem{Chew:1961ev}
G.~F.~Chew and S.~C.~Frautschi,
``Principle of Equivalence for All Strongly Interacting Particles Within the S Matrix Framework,''
Phys. Rev. Lett. \textbf{7}, 394-397 (1961)
doi:10.1103/PhysRevLett.7.394

\bibitem{Kortman:1971zz}
P.~J.~Kortman and R.~B.~Griffiths,
``Density of Zeros on the Lee-Yang Circle for Two Ising Ferromagnets,''
Phys. Rev. Lett. \textbf{27}, 1439-1442 (1971)
doi:10.1103/PhysRevLett.27.1439

\bibitem{Fisher:1978pf}
M.~E.~Fisher,
``Yang-Lee Edge Singularity and phi**3 Field Theory,''
Phys. Rev. Lett. \textbf{40}, 1610-1613 (1978)
doi:10.1103/PhysRevLett.40.1610


\bibitem{Cardy:1985yy}
J.~L.~Cardy,
``Conformal Invariance and the Yang-lee Edge Singularity in Two-dimensions,''
Phys. Rev. Lett. \textbf{54}, 1354-1356 (1985)
doi:10.1103/PhysRevLett.54.1354

\bibitem{Bender:1998ke}
C.~M.~Bender and S.~Boettcher,
``Real spectra in nonHermitian Hamiltonians having PT symmetry,''
Phys. Rev. Lett. \textbf{80}, 5243-5246 (1998)
doi:10.1103/PhysRevLett.80.5243
[arXiv:physics/9712001 [physics]].

\bibitem{Li:2022prn}
W.~Li,
``Null bootstrap for non-Hermitian Hamiltonians,''
Phys. Rev. D \textbf{106}, no.12, 125021 (2022)
doi:10.1103/PhysRevD.106.125021
[arXiv:2202.04334 [hep-th]].


\bibitem{Bender:1999ek}
C.~M.~Bender, K.~A.~Milton and V.~Savage,
``Solution of Schwinger-Dyson equations for PT symmetric quantum field theory,''
Phys. Rev. D \textbf{62}, 085001 (2000)
doi:10.1103/PhysRevD.62.085001
[arXiv:hep-th/9907045 [hep-th]].

\bibitem{Bender:2007nj}
C.~M.~Bender,
``Making sense of non-Hermitian Hamiltonians,''
Rept. Prog. Phys. \textbf{70}, 947 (2007)
doi:10.1088/0034-4885/70/6/R03
[arXiv:hep-th/0703096 [hep-th]].

\bibitem{Bender:2010hf}
C.~M.~Bender and S.~P.~Klevansky,
``Families of particles with different masses in PT-symmetric quantum field theory,''
Phys. Rev. Lett. \textbf{105}, 031601 (2010)
doi:10.1103/PhysRevLett.105.031601
[arXiv:1002.3253 [hep-th]].

\bibitem{r5} 
C.~M.~Bender {\it et al.}, {\it PT Symmetry: in Quantum and
Classical Physics} (World Scientific, Singapore, 2019).

\bibitem{Sibuya}
Y. ~Sibuya, ``Global theory of a second-order linear ordinary differential equation with polynomial coefficient", (Amsterdam: North-Holland 1975).

\bibitem{Li:2024rod}
W.~Li,
``The $\phi^n$ trajectory bootstrap,''
[arXiv:2402.05778 [hep-th]].

\bibitem{Bender:1969si}
C.~M.~Bender and T.~T.~Wu,
``Anharmonic oscillator,''
Phys. Rev. \textbf{184}, 1231-1260 (1969)
doi:10.1103/PhysRev.184.1231

\bibitem{Han:2020bkb}
X.~Han, S.~A.~Hartnoll and J.~Kruthoff,
``Bootstrapping Matrix Quantum Mechanics,''
Phys. Rev. Lett. \textbf{125}, no.4, 041601 (2020)
doi:10.1103/PhysRevLett.125.041601
[arXiv:2004.10212 [hep-th]].

\bibitem{Brezin:1977sv}
E.~Brezin, C.~Itzykson, G.~Parisi and J.~B.~Zuber,
``Planar Diagrams,''
Commun. Math. Phys. \textbf{59}, 35 (1978)
doi:10.1007/BF01614153

\bibitem{Guo:2023qtt}
Y.~Guo and W.~Li,
``Anomalous dimensions of partially conserved higher-spin currents from conformal field theory: Bosonic \ensuremath{\phi}2n theories,''
Phys. Rev. D \textbf{109}, no.2, 025015 (2024)
doi:10.1103/PhysRevD.109.025015
[arXiv:2305.16916 [hep-th]].

\bibitem{El-Showk:2012cjh}
S.~El-Showk, M.~F.~Paulos, D.~Poland, S.~Rychkov, D.~Simmons-Duffin and A.~Vichi,
``Solving the 3D Ising Model with the Conformal Bootstrap,''
Phys. Rev. D \textbf{86}, 025022 (2012)
doi:10.1103/PhysRevD.86.025022
[arXiv:1203.6064 [hep-th]].

\bibitem{El-Showk:2014dwa}
S.~El-Showk, M.~F.~Paulos, D.~Poland, S.~Rychkov, D.~Simmons-Duffin and A.~Vichi,
``Solving the 3d Ising Model with the Conformal Bootstrap II. c-Minimization and Precise Critical Exponents,''
J. Stat. Phys. \textbf{157}, 869 (2014)
doi:10.1007/s10955-014-1042-7
[arXiv:1403.4545 [hep-th]].

\bibitem{Guo:2023gfi}
Y.~Guo and W.~Li,
``Solving anharmonic oscillator with null states: Hamiltonian bootstrap and Dyson-Schwinger equations,''
[arXiv:2305.15992 [hep-th]].

\bibitem{Cornwall:1980zw}
J.~M.~Cornwall,
``Confinement and Chiral Symmetry Breakdown: Estimates of f(pi) and of Effective Quark Masses,''
Phys. Rev. D \textbf{22}, 1452 (1980)
doi:10.1103/PhysRevD.22.1452

\bibitem{Munczek:1983dx}
H.~J.~Munczek and A.~M.~Nemirovsky,
``The Ground State q anti-q Mass Spectrum in QCD,''
Phys. Rev. D \textbf{28}, 181 (1983)
doi:10.1103/PhysRevD.28.181

\bibitem{Krein:1990sf}
G.~Krein, C.~D.~Roberts and A.~G.~Williams,
``On the implications of confinement,''
Int. J. Mod. Phys. A \textbf{7}, 5607-5624 (1992)
doi:10.1142/S0217751X92002544

\bibitem{Li:2017agi}
W.~Li,
``Inverse Bootstrapping Conformal Field Theories,''
JHEP \textbf{01}, 077 (2018)
doi:10.1007/JHEP01(2018)077
[arXiv:1706.04054 [hep-th]].

\bibitem{Li:2017ukc}
W.~Li,
``New method for the conformal bootstrap with OPE truncations,''
[arXiv:1711.09075 [hep-th]].

\bibitem{Shifman:1978bx}
M.~A.~Shifman, A.~I.~Vainshtein and V.~I.~Zakharov,
``QCD and Resonance Physics. Theoretical Foundations,''
Nucl. Phys. B \textbf{147}, 385-447 (1979)
doi:10.1016/0550-3213(79)90022-1

\bibitem{Shifman:1978by}
M.~A.~Shifman, A.~I.~Vainshtein and V.~I.~Zakharov,
``QCD and Resonance Physics: Applications,''
Nucl. Phys. B \textbf{147}, 448-518 (1979)
doi:10.1016/0550-3213(79)90023-3

\bibitem{Poland:2018epd}
D.~Poland, S.~Rychkov and A.~Vichi,
``The Conformal Bootstrap: Theory, Numerical Techniques, and Applications,''
Rev. Mod. Phys. \textbf{91}, 015002 (2019)
doi:10.1103/RevModPhys.91.015002
[arXiv:1805.04405 [hep-th]].

\bibitem{Li:2023tic}
W.~Li,
``Easy bootstrap for the 3D Ising model,''
[arXiv:2312.07866 [hep-th]].


\end{thebibliography}
\end{document}